\documentclass[sigplan,screen]{acmart}

\acmConference[PriSC'24]{ACM Workshop on Principles of Secure Compilation}{January 17--19, 2024}{London, United Kingdom}
\acmYear{2024}
\acmISBN{} 
\acmDOI{} 
\startPage{1}

\setcopyright{none}

\usepackage{booktabs}   
\usepackage{subcaption} 
\usepackage{listings}
\usepackage{wrapfig}
\usepackage{setspace}
\usepackage{newunicodechar}
\usepackage{placeins}
\usepackage{xspace}
\newunicodechar{≠}{=\llap{/}}
\newunicodechar{≔}{:\raisebox{-0.18ex}[\height][\depth]{=}}
\newunicodechar{≤}{\rlap{\begingroup \fontsize{13pt}{13pt}\selectfont \raisebox{-0.58ex}[\height][\depth]{-} \endgroup}\shifttext{0.15ex}{\raisebox{0.18ex}[\height][\depth]{<}}}
\newunicodechar{≥}{\rlap{\begingroup \fontsize{13pt}{13pt}\selectfont \raisebox{-0.58ex}[\height][\depth]{-} \endgroup}\shifttext{0.15ex}{\raisebox{0.18ex}[\height][\depth]{>}}}
\newunicodechar{∧}{\shifttext{1ex}{\begingroup \fontsize{8pt}{8pt}\selectfont {\shifttext{0.4ex}{/}\shifttext{-0.4ex}{\textbackslash}} \endgroup}}
\newunicodechar{∨}{\shifttext{1ex}{\begingroup \fontsize{8pt}{8pt}\selectfont {\shifttext{0.4ex}{\textbackslash}\shifttext{-0.4ex}{/}} \endgroup}}

\makeatletter
\newcommand*{\shifttext}[2]{%
	\settowidth{\@tempdima}{#2}%
	\makebox[\@tempdima]{\hspace*{#1}#2}%
}
\makeatother

\usepackage[leftmargin=3em,vskip=0em]{quoting}
  
\definecolor{light-gray}{gray}{0.87}

\definecolor{gray}{gray}{0.75}

\definecolor{drk-gray}{RGB}{169,169,169}

\definecolor{light-purple}{RGB}{229,204,255}

\definecolor{light-yellow}{RGB}{255,228,181}

\definecolor{light-blue}{RGB}{189,215,238}

\definecolor{light-green}{RGB}{152,251,152}

\definecolor{light-red}{RGB}{255,204,204}

\definecolor{rred}{RGB}{255,153,153}

\definecolor{light-pink}{RGB}{255,204,255}

\definecolor{light-orange}{RGB}{255,204,153}

\definecolor{neon-blue}{RGB}{153,255,255}

\definecolor{neon-yellow}{RGB}{255,255,153}

\newcommand{\ttt}[1]{\textup{\texttt{\small{#1}}}}

\newcommand{\qm}{\ttt{?}}

\usepackage{pdfpages}
\usepackage{multirow}
\usepackage{multicol}
\usepackage{array} 
\usepackage{savesym} 
\savesymbol{newlist}
\savesymbol{renewlist}
\usepackage{enumitem}
\restoresymbol{enumi}{newlist}
\restoresymbol{enumi}{renewlist}

\usepackage{tikz}

\definecolor{legend_green}{HTML}{33A02C}
\definecolor{legend_purple}{HTML}{6A3D9A}
\definecolor{legend_red}{HTML}{DE2D26}
\definecolor{legend_yellow}{HTML}{FFD700}
\definecolor{legend_green_light}{HTML}{c8e4c4}
\definecolor{legend_purple_light}{HTML}{c4c4e8}

\usepackage{algorithmicx}
\usepackage{algpseudocode,algorithm,algorithmicx}

\algrenewcommand\algorithmicrequire{\textbf{Precondition:}}
\algrenewcommand\algorithmicensure{\textbf{Postcondition:}}

\newlength\myindent
\begin{document}

\title[Gradual Verification for Smart Contracts]{Gradual Verification for Smart Contracts}         


\author{Haojia Sun}
\orcid{nnnn-nnnn-nnnn-nnnn}             
\affiliation{
  \institution{Shanghai Jiao Tong University}   
}
\makeatletter
\def\@shortauthors{Sun, Singh, Ramos-Dávila, Aldrich and DiVincenzo}
\makeatother

\author{Kunal Singh}
\orcid{nnnn-nnnn-nnnn-nnnn}             
\affiliation{
  \institution{Carnegie Mellon University}   
}

\author{Jan-Paul Ramos-Dávila}
\orcid{nnnn-nnnn-nnnn-nnnn}             
\affiliation{
  \institution{Cornell University}   
}

\author{Jonathan Aldrich}
\orcid{nnnn-nnnn-nnnn-nnnn}             
\affiliation{
  \institution{Carnegie Mellon University} 
}

\author{Jenna DiVincenzo}
\orcid{nnnn-nnnn-nnnn-nnnn}             
\affiliation{
  \institution{Purdue University}   
}

\begin{abstract}
Blockchains facilitate secure resource transactions through smart contracts, yet these digital agreements are prone to vulnerabilities, particularly when interacting with external contracts, leading to substantial monetary losses. Traditional verification techniques fall short in providing comprehensive security assurances, especially against re-entrancy attacks, due to the unavailable implementations of external contracts. This paper introduces an incremental approach: \textit{gradual verification}. We combine static and dynamic verification techniques to enhance security, guarantee soundness and flexibility, and optimize resource usage in smart contract interactions. By implementing a prototype for gradually verifying Algorand smart contracts via the pyTEAL language, we demonstrate the effectiveness of our approach, contributing to the safe and efficient execution of smart contracts.
\end{abstract}



\maketitle

\section{Introduction}
{Smart contracts, self-executing programs on blockchains like Algorand and Ethereum, facilitate secure resource transactions among mutually untrusted parties~\cite{smartContractDef, 2018EthereumSemantic}. Despite their potential, bugs in smart contracts come with serious consequences, such as substantial monetary loss~\cite{financialLoss}. Therefore, it is important to ensure the correctness of smart contracts.}

{Verification is particularly challenging when one smart contract calls another, which is external. The external contract may not be verified and may break the assumptions its caller makes, particularly in the case of re-entrant calls like those at the root of the DAO attack on Ethereum~\cite{DAOattack}.}


{Recent verification approaches have been introduced to tackle this problem. \citet{Petermuller, hajdu2020solc, hildenbrandt2018kevm}, and \citet{kalra2018zeus} introduce static verification techniques capable of verifying the functional properties of Ethereum smart contracts, even when they involve calls to external contracts. These approaches provide strong guarantees, but they burden developers by demanding exhaustive and meticulous specifications. \citet{shyamasundar, sereum2018}, and \citet{li2020} introduce approaches that dynamically monitor and enforce user-specified invariants in smart contracts that call external contracts. While these dynamic techniques offer flexibility and ease of use, they can only uncover vulnerabilities in executed program paths at run time. 
Moreover, run-time checks incur substantial transaction fees on blockchains, increasing the monetary cost of executing smart contracts.}

{Given these constraints, gradual verification \cite{bader2018gradual, wise2020gradual} emerges as a compelling solution for securing smart contracts. Gradual verification supports partial specifications and incremental verification of code by applying static verification where possible and dynamic verification where necessary. This gives developers control over the trade-offs between static and dynamic verification. They can write more static specifications and get stronger guarantees and less run time overhead; or, they can write fewer specifications---saving on human effort ---and rely more on run-time checking and its cost. The spectrum of trade-offs is formally guaranteed \cite{bader2018gradual, wise2020gradual}. We present a prototype for gradually verifying smart contracts via TEAL, the programming language used for creating Algorand smart contracts. By extending a gradual verification architecture from prior work \cite{divincenzo2022gradual,wise2020gradual}, we can provide the following benefits to smart contracts and their developers:

\paragraph{Protecting Against Unverified Code \& Re-entrancy.}
{Traditional static verification techniques are unsound in the face of unverified code and arbitrary re-entrancy.
In contrast, gradual verification soundly guards against undefined behavior in unverified code and from arbitrary re-entrancy by run-time checking pre- and postconditions---that may be partially specified---on current and external contracts.
Furthermore, for critical sections of available code, static verification can be employed proactively to identify potential issues. 

\paragraph{Balancing Precision and Flexibility.}
{The incremental approach to verification supported by gradual verification is well-suited for the fast-paced and ever-changing blockchain environment. It also makes verification more accessible to novice developers of smart contracts. Gradual verification suppresses static verification errors from missing specifications allowing developers to specify only the software components and properties they care about! Only true static and dynamic errors are reported, allowing bugs and vulnerabilities in smart contracts and their specifications to be detected earlier than static or dynamic verification alone \cite{divincenzo2022gradual}.

\paragraph{Reducing Run-time Overhead.}
{The transaction fees and computational resources required for executing smart contracts can accumulate quickly, like the Ethereum gas fee, which increases with dynamic verification alone. Gradual verification minimizes this cost by minimizing run-time checks with statically available information---when a proof obligation has been statically verified, no run-time check is generated for it. Therefore, developers may choose to spend more time incrementally specifying their software to reduce dynamic checking overhead.

\section{Architecture: Gradual pyTEAL}
{We present a pipeline (see Fig. \ref{fig:pipeline}) implementing the gradual verification of pyTEAL smart contracts, part of the Algorand platform. pyTEAL contracts are written in a domain-specific subset of Python, to which we added pre- and post-condition assertions. Our approach converts pyTEAL programs to Gradual C0, the only previously implemented gradual verifier~\cite{divincenzo2022gradual}. Internally, Gradual C0 uses a variant of the Viper toolchain~\cite{Petermuller} to statically verify functions, treating partial specifications optimistically. The modified Viper back-end generates run-time checks wherever static verification could not assure specifications; these run-time checks are added to the original smart contract before compilation to the Algorand platform.
}


\setlength{\intextsep}{4pt}
\setlength{\belowcaptionskip}{-3pt}
\begin{figure}[H]
    \includegraphics[width=0.49\textwidth]{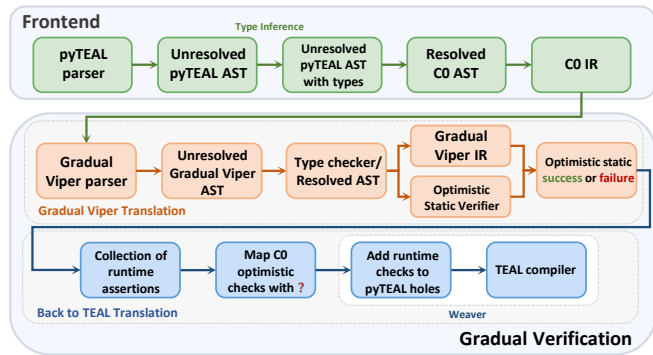}
    \caption{Gradual pyTEAL verification pipeline}
    \label{fig:pipeline}
\end{figure}

\paragraph{Front-end.} 
{We construct a parser using the FastParse Scala library to parse pyTEAL source code with specifications into an unresolved pyTEAL Abstract Syntax Tree (AST). 
Since Python supports dynamic typing but we are translating to a typed intermediate language (C0), we use type inference to assign static types to all variables in the pyTEAL AST, resulting in an unresolved pyTEAL AST with inferred types. We perform type checking and other well-formedness checks, then translate this "resolved AST" to C0. The C0 AST makes some operations more explicit, aiding verification and connecting to the intermediate representation supported by Gradual C0.
}

\paragraph{Static Verification.}

{During static verification, the verifier \textit{optimistically} interprets \emph{imprecise formulas} to satisfy a contract (denoted by $\qm$), with a promise that these holes in the specification will be dealt with at run time. When optimistic static verification succeeds, a set of run-time checks are produced that must be executed at run time for soundness.}

\paragraph{Dynamic Verification.}
{We then identify the locations where run-time checks should be inserted, based on the placement of $\qm$ embedded in our pyTEAL specifications. We developed the Weaver module to perform this step and encode the checks in the original pyTEAL program at the identified locations. 
Finally, the pyTEAL program with run-time checks is transformed into an executable contract with the pyTEAL compiler, which is subsequently executed on the Algorand platform. In the event of run-time check failures, corresponding error messages are reported.
In practice, run-time checks may contain simple logical expressions,
accessibility predicates denoting ownership of contract state,
and complex predicates implemented as recursive boolean functions.}


\subsection{Example}
\begin{figure}[H]
\begin{lstlisting}[
    aboveskip=1pt, 
    belowskip=1pt, 
    xleftmargin=1.5em,
    language=Python, 
    basicstyle=\ttfamily\small, 
    breaklines=true,
    keywordstyle=\bfseries\color{red!60!yellow},
    morekeywords={},
    emph={self}, % 
    emphstyle=\bfseries\color{Rhodamine}, 
    commentstyle=\itshape\color{blue!70!white}, 
    columns=flexible,
    numbers=left,
    numbersep=1em, 
    numberstyle=\footnotesize
]  
#@ global Count;
@router.method
def sell(quantity: abi.Uint64):
   #@ requires ? and quantity>=0 and acc(Count);
   #@ ensures Count >=0;
   scratchCount = ScratchVar(TealType.uint64)
   return Seq(
      scratchCount.store(App.globalGet(
           Bytes("Count"))),
      App.globalPut(Bytes("Count"), scratchCount.load() - quantity.get())
   )
\end{lstlisting}
\caption{A gradually verified Algorand smart contract}
\label{fig:algorand-example}
\end{figure}

{In Figure \ref{fig:algorand-example}, a segment of an Algorand smart contract depicts a selling transaction, where a specified \ttt{quantity} is subtracted from a global state variable \ttt{Count}. In Algorand smart contracts, global state refers to storage that persists across different contract calls and is accessible to all instances of the contract. The \ttt{//@requires} annotation in the code ensures that \ttt{quantity} is non-negative and verifies access to the \ttt{Count} variable.  A $\qm$ indicates that this specification is partial. By specifying access permissions, the contract ensures that functions interact with the global state as intended, safeguarding against unauthorized modifications and security threats, thus maintaining the integrity of the state management.

Statically, the \ttt{//@ensures} annotation specifies a postcondition, asserting that \ttt{Count} remains non-negative after the transaction. This postcondition cannot be proven statically because an additional precondition, \ttt{quantity <= Count}, would be required. However, the verifier optimistically assumes this property can be derived from the $\qm$, and a run-time check is added to make sure the property is obeyed at run time.  By requiring the developer to explicitly specify data types, access controls, and preconditions/postconditions, the gradual verification architecture ensures that these specifications can be optimistically statically checked before the run time, while ensuring run-time soundness.}


\section{Acknowledgements}
The authors thank the PriSC reviewers for their helpful comments. This research was supported by the  \grantsponsor{NSF}{National Science Foundation}{https://www.nsf.gov/} under Grant
  No. \grantnum[https://www.nsf.gov/awardsearch/showAward?AWD_ID=1901033]{NSF}{CCF-1901033} and by a \grantsponsor{Google}{Google PhD Fellowship}{https://research.google/outreach/phd-fellowship/}.  Any opinions, findings, and conclusions or recommendations expressed in this material are those of the authors and do not necessarily reflect the views of the National Science Foundation or Google.

\bibliography{gradual_verification_for_smart_contracts}


\begin{thebibliography}{14}


\ifx \showCODEN    \undefined \def \showCODEN     #1{\unskip}     \fi
\ifx \showDOI      \undefined \def \showDOI       #1{#1}\fi
\ifx \showISBNx    \undefined \def \showISBNx     #1{\unskip}     \fi
\ifx \showISBNxiii \undefined \def \showISBNxiii  #1{\unskip}     \fi
\ifx \showISSN     \undefined \def \showISSN      #1{\unskip}     \fi
\ifx \showLCCN     \undefined \def \showLCCN      #1{\unskip}     \fi
\ifx \shownote     \undefined \def \shownote      #1{#1}          \fi
\ifx \showarticletitle \undefined \def \showarticletitle #1{#1}   \fi
\ifx \showURL      \undefined \def \showURL       {\relax}        \fi
\providecommand\bibfield[2]{#2}
\providecommand\bibinfo[2]{#2}
\providecommand\natexlab[1]{#1}
\providecommand\showeprint[2][]{arXiv:#2}

\bibitem[\protect\citeauthoryear{Bader, Aldrich, and Tanter}{Bader et~al\mbox{.}}{2018}]%
        {bader2018gradual}
\bibfield{author}{\bibinfo{person}{Johannes Bader}, \bibinfo{person}{Jonathan Aldrich}, {and} \bibinfo{person}{{\'E}ric Tanter}.} \bibinfo{year}{2018}\natexlab{}.
\newblock \showarticletitle{Gradual program verification}. In \bibinfo{booktitle}{\emph{Verification, Model Checking, and Abstract Interpretation: 19th International Conference, VMCAI 2018, Los Angeles, CA, USA, January 7-9, 2018, Proceedings 19}}. Springer, \bibinfo{pages}{25--46}.
\newblock


\bibitem[\protect\citeauthoryear{Br{\"a}m, Eilers, M{\"u}ller, Sierra, and Summers}{Br{\"a}m et~al\mbox{.}}{2021}]%
        {Petermuller}
\bibfield{author}{\bibinfo{person}{Christian Br{\"a}m}, \bibinfo{person}{Marco Eilers}, \bibinfo{person}{Peter M{\"u}ller}, \bibinfo{person}{Robin Sierra}, {and} \bibinfo{person}{Alexander~J Summers}.} \bibinfo{year}{2021}\natexlab{}.
\newblock \showarticletitle{Rich specifications for Ethereum smart contract verification}.
\newblock \bibinfo{journal}{\emph{Proceedings of the ACM on Programming Languages}} \bibinfo{volume}{5}, \bibinfo{number}{OOPSLA} (\bibinfo{year}{2021}), \bibinfo{pages}{1--30}.
\newblock


\bibitem[\protect\citeauthoryear{DiVincenzo, McCormack, Gouni, Gorenburg, Zhang, Zimmerman, Sunshine, Tanter, and Aldrich}{DiVincenzo et~al\mbox{.}}{2022}]%
        {divincenzo2022gradual}
\bibfield{author}{\bibinfo{person}{Jenna DiVincenzo}, \bibinfo{person}{Ian McCormack}, \bibinfo{person}{Hemant Gouni}, \bibinfo{person}{Jacob Gorenburg}, \bibinfo{person}{Mona Zhang}, \bibinfo{person}{Conrad Zimmerman}, \bibinfo{person}{Joshua Sunshine}, \bibinfo{person}{{\'E}ric Tanter}, {and} \bibinfo{person}{Jonathan Aldrich}.} \bibinfo{year}{2022}\natexlab{}.
\newblock \showarticletitle{Gradual C0: Symbolic Execution for Efficient Gradual Verification}.
\newblock \bibinfo{journal}{\emph{arXiv preprint arXiv:2210.02428}} (\bibinfo{year}{2022}).
\newblock


\bibitem[\protect\citeauthoryear{Grishchenko, Maffei, and Schneidewind}{Grishchenko et~al\mbox{.}}{2018}]%
        {2018EthereumSemantic}
\bibfield{author}{\bibinfo{person}{Ilya Grishchenko}, \bibinfo{person}{Matteo Maffei}, {and} \bibinfo{person}{Clara Schneidewind}.} \bibinfo{year}{2018}\natexlab{}.
\newblock \showarticletitle{A semantic framework for the security analysis of ethereum smart contracts}. In \bibinfo{booktitle}{\emph{Principles of Security and Trust: 7th International Conference, POST 2018, Held as Part of the European Joint Conferences on Theory and Practice of Software, ETAPS 2018, Thessaloniki, Greece, April 14-20, 2018, Proceedings 7}}. Springer, \bibinfo{pages}{243--269}.
\newblock


\bibitem[\protect\citeauthoryear{Hajdu and Jovanovi{\'c}}{Hajdu and Jovanovi{\'c}}{2020}]%
        {hajdu2020solc}
\bibfield{author}{\bibinfo{person}{{\'A}kos Hajdu} {and} \bibinfo{person}{Dejan Jovanovi{\'c}}.} \bibinfo{year}{2020}\natexlab{}.
\newblock \showarticletitle{solc-verify: A modular verifier for solidity smart contracts}. In \bibinfo{booktitle}{\emph{Verified Software. Theories, Tools, and Experiments: 11th International Conference, VSTTE 2019, New York City, NY, USA, July 13--14, 2019, Revised Selected Papers 11}}. Springer, \bibinfo{pages}{161--179}.
\newblock


\bibitem[\protect\citeauthoryear{Hildenbrandt, Saxena, Rodrigues, Zhu, Daian, Guth, Moore, Park, Zhang, Stefanescu, et~al\mbox{.}}{Hildenbrandt et~al\mbox{.}}{2018}]%
        {hildenbrandt2018kevm}
\bibfield{author}{\bibinfo{person}{Everett Hildenbrandt}, \bibinfo{person}{Manasvi Saxena}, \bibinfo{person}{Nishant Rodrigues}, \bibinfo{person}{Xiaoran Zhu}, \bibinfo{person}{Philip Daian}, \bibinfo{person}{Dwight Guth}, \bibinfo{person}{Brandon Moore}, \bibinfo{person}{Daejun Park}, \bibinfo{person}{Yi Zhang}, \bibinfo{person}{Andrei Stefanescu}, {et~al\mbox{.}}} \bibinfo{year}{2018}\natexlab{}.
\newblock \showarticletitle{Kevm: A complete formal semantics of the ethereum virtual machine}. In \bibinfo{booktitle}{\emph{2018 IEEE 31st Computer Security Foundations Symposium (CSF)}}. IEEE, \bibinfo{pages}{204--217}.
\newblock


\bibitem[\protect\citeauthoryear{Kalra, Goel, Dhawan, and Sharma}{Kalra et~al\mbox{.}}{2018}]%
        {kalra2018zeus}
\bibfield{author}{\bibinfo{person}{Sukrit Kalra}, \bibinfo{person}{Seep Goel}, \bibinfo{person}{Mohan Dhawan}, {and} \bibinfo{person}{Subodh Sharma}.} \bibinfo{year}{2018}\natexlab{}.
\newblock \showarticletitle{Zeus: analyzing safety of smart contracts.}. In \bibinfo{booktitle}{\emph{Ndss}}. \bibinfo{pages}{1--12}.
\newblock


\bibitem[\protect\citeauthoryear{Li, Choi, and Long}{Li et~al\mbox{.}}{2020}]%
        {li2020}
\bibfield{author}{\bibinfo{person}{Ao Li}, \bibinfo{person}{Jemin~Andrew Choi}, {and} \bibinfo{person}{Fan Long}.} \bibinfo{year}{2020}\natexlab{}.
\newblock \showarticletitle{Securing smart contract with runtime validation}. In \bibinfo{booktitle}{\emph{Proceedings of the 41st ACM SIGPLAN Conference on Programming Language Design and Implementation}}. \bibinfo{pages}{438--453}.
\newblock


\bibitem[\protect\citeauthoryear{Mehar, Shier, Giambattista, Gong, Fletcher, Sanayhie, Kim, and Laskowski}{Mehar et~al\mbox{.}}{2019}]%
        {DAOattack}
\bibfield{author}{\bibinfo{person}{Muhammad~Izhar Mehar}, \bibinfo{person}{Charles~Louis Shier}, \bibinfo{person}{Alana Giambattista}, \bibinfo{person}{Elgar Gong}, \bibinfo{person}{Gabrielle Fletcher}, \bibinfo{person}{Ryan Sanayhie}, \bibinfo{person}{Henry~M Kim}, {and} \bibinfo{person}{Marek Laskowski}.} \bibinfo{year}{2019}\natexlab{}.
\newblock \showarticletitle{Understanding a revolutionary and flawed grand experiment in blockchain: the DAO attack}.
\newblock \bibinfo{journal}{\emph{Journal of Cases on Information Technology (JCIT)}} \bibinfo{volume}{21}, \bibinfo{number}{1} (\bibinfo{year}{2019}), \bibinfo{pages}{19--32}.
\newblock


\bibitem[\protect\citeauthoryear{Ndiaye and Konate}{Ndiaye and Konate}{2021}]%
        {financialLoss}
\bibfield{author}{\bibinfo{person}{Malaw Ndiaye} {and} \bibinfo{person}{Pr~Karim Konate}.} \bibinfo{year}{2021}\natexlab{}.
\newblock \showarticletitle{Cryptocurrency crime: Behaviors of malicious smart contracts in blockchain}. In \bibinfo{booktitle}{\emph{2021 International Symposium on Networks, Computers and Communications (ISNCC)}}. IEEE, \bibinfo{pages}{1--8}.
\newblock


\bibitem[\protect\citeauthoryear{Rodler, Li, Karame, and Davi}{Rodler et~al\mbox{.}}{2018}]%
        {sereum2018}
\bibfield{author}{\bibinfo{person}{Michael Rodler}, \bibinfo{person}{Wenting Li}, \bibinfo{person}{Ghassan~O Karame}, {and} \bibinfo{person}{Lucas Davi}.} \bibinfo{year}{2018}\natexlab{}.
\newblock \showarticletitle{Sereum: Protecting existing smart contracts against re-entrancy attacks}.
\newblock \bibinfo{journal}{\emph{arXiv preprint arXiv:1812.05934}} (\bibinfo{year}{2018}).
\newblock


\bibitem[\protect\citeauthoryear{Shyamasundar}{Shyamasundar}{2022}]%
        {shyamasundar}
\bibfield{author}{\bibinfo{person}{RK Shyamasundar}.} \bibinfo{year}{2022}\natexlab{}.
\newblock \showarticletitle{A Framework of Runtime Monitoring for Correct Execution of Smart Contracts}. In \bibinfo{booktitle}{\emph{International Conference on Blockchain}}. Springer, \bibinfo{pages}{92--116}.
\newblock


\bibitem[\protect\citeauthoryear{Szabo}{Szabo}{1997}]%
        {smartContractDef}
\bibfield{author}{\bibinfo{person}{Nick Szabo}.} \bibinfo{year}{1997}\natexlab{}.
\newblock \showarticletitle{Formalizing and securing relationships on public networks}.
\newblock \bibinfo{journal}{\emph{First monday}} (\bibinfo{year}{1997}).
\newblock


\bibitem[\protect\citeauthoryear{Wise, Bader, Wong, Aldrich, Tanter, and Sunshine}{Wise et~al\mbox{.}}{2020}]%
        {wise2020gradual}
\bibfield{author}{\bibinfo{person}{Jenna Wise}, \bibinfo{person}{Johannes Bader}, \bibinfo{person}{Cameron Wong}, \bibinfo{person}{Jonathan Aldrich}, \bibinfo{person}{{\'E}ric Tanter}, {and} \bibinfo{person}{Joshua Sunshine}.} \bibinfo{year}{2020}\natexlab{}.
\newblock \showarticletitle{Gradual verification of recursive heap data structures}.
\newblock \bibinfo{journal}{\emph{Proceedings of the ACM on Programming Languages}} \bibinfo{volume}{4}, \bibinfo{number}{OOPSLA} (\bibinfo{year}{2020}), \bibinfo{pages}{1--28}.
\newblock


\end{thebibliography}



\end{document}